\documentclass[a4paper, onecolumn, 10pt]{article}

\usepackage[center,width=152mm,height=235mm]{crop}

\usepackage[english]{babel}
\usepackage[latin1]{inputenc}
\usepackage[T1]{fontenc}
\usepackage{selinput}
\SelectInputMappings{%
  aacute={á},
  ntilde={ñ},
  Euro={€}
}

\usepackage{float} % For controlling figure positions
\usepackage{amsthm, amsmath, amsfonts, amssymb}
\usepackage{fancybox}
\usepackage{color}
\usepackage{url}
\usepackage{appendix}
\usepackage[authoryear]{natbib}
\usepackage{graphicx}
\graphicspath{ {./img/} }

\usepackage[draft, colorinlistoftodos]{todonotes} % Use disable instead of draft to hide todo notes

\usepackage[switch]{lineno} % Activate using \linenumbers

\newcommand{\numReps}{200} % Number of repetitions
\newcommand{\stabilTime}{2000} % Stabilization time
\newcommand{\simTime}{5000} % Simulation time
\newcommand{\numEpsilons}{25} % Number of epsilons to be tried
\newcommand{\bioThreshold}{0.01} % Threshold to consider extinction
\newcommand{\windowWidth}{0.05} % Window for the random non-diagonal competition parameters

\usepackage{authblk}
\title{Neutral competition boosts chaos in food webs}
\author[1]{Pablo Rodríguez-Sánchez \thanks{\texttt{pablo.rodriguezsanchez@wur.nl}}}
\author[1]{Egbert H. van Nes \thanks{\texttt{egbert.vannes@wur.nl}}}
\author[1]{Marten Scheffer \thanks{\texttt{marten.scheffer@wur.nl}}}

\affil[1]{Department of Aquatic Ecology, Wageningen University, The Netherlands}

\begin{document}

\maketitle

%% Abstract
\begin{abstract}
\label{sec:Abstract}
Similarity of competitors has been proposed to facilitate coexistence of species because it slows down competitive exclusion, thus making it easier for equalizing mechanisms to maintain diverse communities. On the other hand, chaos can promote coexistence of species. Here we link these two previously unrelated findings, by analyzing the dynamics of food web models. We show that near-neutrality of competition of prey, in the presence of predators, increases the chance of developing chaotic dynamics. Moreover we confirm that this results in a higher biodiversity. Our results suggest that near-neutrality may promote biodiversity in two ways: through reducing the rates of competitive displacement and through promoting non-equilibrium dynamics.

\end{abstract}

%% Main body of the paper
\newpage
\section{Background}
\label{sec:Background}
Ever since Darwin, the idea that species must be sufficiently different to coexist is deeply rooted in biological thinking. Indeed, the principles of limiting similarity \citep{MacArthur} and competitive exclusion \citep{Hardin1960, Armstrong1980} are the cornerstones of ecological theory. Nevertheless, natural communities (such as plankton communities \citep{Hutchinson1961}), often harbor far more species that may be explained from niche separation, inspiring G. Evelyn Hutchinson \citeyearpar{Hutchinson} to ask the simple but fundamental question \textit{"why are there so many kinds of animals?"}. Since then many mechanisms have been suggested that may help similar species to coexist. As Hutchinson \citeyearpar{Hutchinson1961} already proposed himself, fluctuations in conditions may prevent populations to reach equilibrium at which species would be outcompeted. Also, natural enemies including pests and parasites tend to attack the abundant species more than rare species, and such a \textit{"kill the winner"} \citep{Winter2010} mechanism promotes diversity by preventing one species to become dominant.

In the extensive literature on potential mechanisms that could prevent competitive exclusion there are two relatively new ideas that have created some controversy: neutrality and chaos. The neutral theory of biodiversity introduced by Hubbell \citeyearpar{Hubbell2001} proposes that species that are entirely equivalent can coexist because none is able to outcompete the other. The concept of equivalent species has met skepticism as it is incompatible with the idea that all species are different. However, it turns out that also \textit{"near-neutral"} competitors can coexist in models of competition and evolution \citep{Scheffer2006, Scheffer2018}. Support for such near-neutrality has been found in a wide range of communities \citep{Vergnon2013, Segura2013}. The second controversial mechanism that may prevent competitive exclusion is \textit{"super-saturated coexistence"} in communities that display chaotic dynamics \citep{Huisman1999}. This is in a sense analogous to the prevention of competitive exclusion in fluctuating environments, except that deterministic chaos is internally driven. Although there has been much debate about the question whether chaotic dynamics plays an important role in ecosystems \citep{Berryman1989, Scheffer1991, Schippers2001}, several studies support the idea that chaos can be an essential ingredient of natural dynamics \citep{Huisman1999, Beninca2008, Beninca}.

Intuitively, it seems not likely that chaos and neutrality can be related, as fully neutral ecosystems can't be chaotic. However, natural ecosystems are never perfectly neutral, and predators may have a preference for different species. In the present work, we used a multi-species food-web model to explore the effect of near-neutrality of prey on the probability of developing chaotic dynamics. We found a surprising link between both ideas: the closer to neutrality the competition is, the higher the chances of developing chaotic dynamics. Additionally, our results confirmed that there is a positive relation between cyclic or chaotic dynamics and the number of coexisiting species.

\section{Methods}
\label{sec:Methods}

\subsection{Model description}
\label{subsec:Model}

We focused our attention on food webs with two trophic levels, competing prey and predators. The predators (consumers) have a differentiated preference of different prey species.

% Description of the dynamics
The dynamics were modelled using the Rosenzweig-MacArthur predator-prey model \citep{Rosenzweig1963}, generalized to a higher number of species \citep{Scheffer2004}. Our model contains $n_P$ prey species and $n_C$ predator species. The prey’s populations are under the influence of both intra and interspecific competition, whose intensities are defined by the competition matrix $A$. The relative preference that predators have for each prey is defined by the predation matrix $S$. Prey immigration from neighboring areas has been added to the classical model in order to avoid unrealistic dynamics, such as heteroclinic orbits giving rise to long-stretched cycles with near extinctions \citep{Scheffer2004}. In mathematical notation, the system reads:
\begin{eqnarray}
\label{eq:SystemUnderStudy}
	\begin{cases}
	\frac{dP_i}{dt} =  r_i(P) P_i  - \sum_{j = 1}^{n_C} g_j(P) P_i S_{ji} C_j + f & : i = 1..n_P
	\\
	\frac{dC_j}{dt} = - l C_j +  e \sum_{i = 1}^{n_P} g_j(P) P_i S_{ji} C_j  & : j = 1..n_C
	\end{cases}
\end{eqnarray}
where $P_i(t)$ represents the biomass of prey species $i$ at time $t$ and $C_j(t)$ the biomass of predator species $j$ at time $t$. The symbol $P$ is used as a shorthand for the vector $(P_1(t), P_2(t), ..., P_{n_P}(t))$. The auxiliary functions $r_i(P)$ and $g_j(P)$ (see equations \eqref{eq:LogisticGenerator} and \eqref{eq:HollingGenerator}) have been respectively chosen to generalize the logistic growth and the Holling type II saturation functional response \citep{Edelstein-Keshet} to a multispecies system when inserted into equation \eqref{eq:SystemUnderStudy}.
\begin{equation}
\label{eq:LogisticGenerator}
	r_i(P) = r \left( 1 - \frac{1}{K} \sum_{k=1}^{n_P} A_{ik} P_k \right)
\end{equation}
\begin{equation}
\label{eq:HollingGenerator}
	g_j(P) = \frac{g}{\sum_{i=1}^{n_P} S_{ji} P_i + H}
\end{equation}
For details about the parameters used, please refer to subsection \ref{subsec:Parameterization}.

\subsection{Parameterization}
\label{subsec:Parameterization}
We parameterized our model as a freshwater plankton system based on Dakos' model \citep{Dakos2009b}. Unlike Dakos, who uses seasonally changing parameters, our parameters were assumed to be independent of time (see table \ref{tab:Parameters}).

\begin{table}[H]
	\begin{center}
		\resizebox{\columnwidth}{!}{
		\begin{tabular}{cccc}
			\hline
			\textbf{Symbol} & \textbf{Interpretation} & \textbf{Value} & \textbf{Units} \\
			\hline
			$r$ & Maximum growth rate & $0.50$ & $d^{-1}$ \\
		    \hline
			$K$ & Carrying capacity & $10.00$ & $ mg \ l^{-1} $ \\
			\hline
			$g$ & Predation rate & $0.40$ & $d^{-1}$\\
			\hline
			$f$ & Immigration rate & $10^{-5}$ & $mg \ l^{-1} \ d^{-1}$\\
			\hline
			$e$ & Assimilation efficiency & $0.60$ & $1$\\
			\hline
			$H$ & Saturation constant & $2.00$ & $ mg \ l^{-1} $\\
		    \hline
			$l$ & Predator's loss rate & $0.15$ & $d^{-1}$\\
			\hline
		    $S$ & $ n_C \times n_P $ predator preference matrix & See section \ref{subsubsec:CompetitionParameter} & $1$\\
		    \hline
   		    $A$ & $ n_P \times n_P $ competition matrix & See section \ref{subsubsec:CompetitionParameter} & $1$\\
		    \hline
		\end{tabular}}
	\end{center}
	\caption{Values and meanings of the parameters used in our numerical experiment. The elements of the predation ($S$) and competition ($A$) matrices are drawn from probability distributions described in subsection \ref{subsubsec:CompetitionParameter}.}
	\label{tab:Parameters}
\end{table}

\subsubsection{Competition and predation matrices}
\label{subsubsec:CompetitionParameter}

\begin{figure}[H]
	\begin{center}
		\includegraphics[width=1\columnwidth]{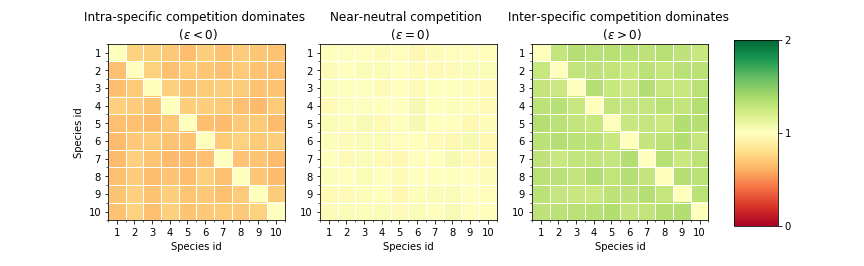}
	\end{center}
	\caption{The competition matrix on the left is a clear case of dominant intraspecific competition. The central one represents a case of near-neutral competition. The matrix in the right panel shows a case of dominant interspecific competition. The difference between them is the relative size of the non-diagonal elements respective of the diagonal ones. This property of the competition matrices is controlled by the competition parameter $\epsilon$.}
	\label{fig:CompetitionParameter}
\end{figure}

Our main purpose is to analyze the effect of different competition strengths on the long term dynamics exhibited. For this, we introduce the competition parameter $\epsilon$ to build a competition matrix $A$, whose non-diagonal terms are drawn from a uniform distribution centered at $1+\epsilon$ and with a given width (here we chose $ w = \windowWidth$). The diagonal terms are by definition 1. Defined this way, the parameter $\epsilon$ allows us to move continuously from strong intraspecific ($ \epsilon < 0$) to strong interspecific competition ($ \epsilon > 0$), meeting neutral-on-average competition at $\epsilon = 0$. For the rest of this paper, we will call ecosystems with $ \epsilon = 0$ \textit{near-neutral} (see figure \ref{fig:CompetitionParameter}).

Regarding the predation matrix $S$ , we follow \citet{Dakos2009b} and draw each of its coefficients from a uniform probability distribution bounded between $0$ and $1$.

\subsection{Numerical experiments}
\label{subsec:NumericalExperiment}

\begin{figure}
	\begin{center}
		\includegraphics[width=1\columnwidth]{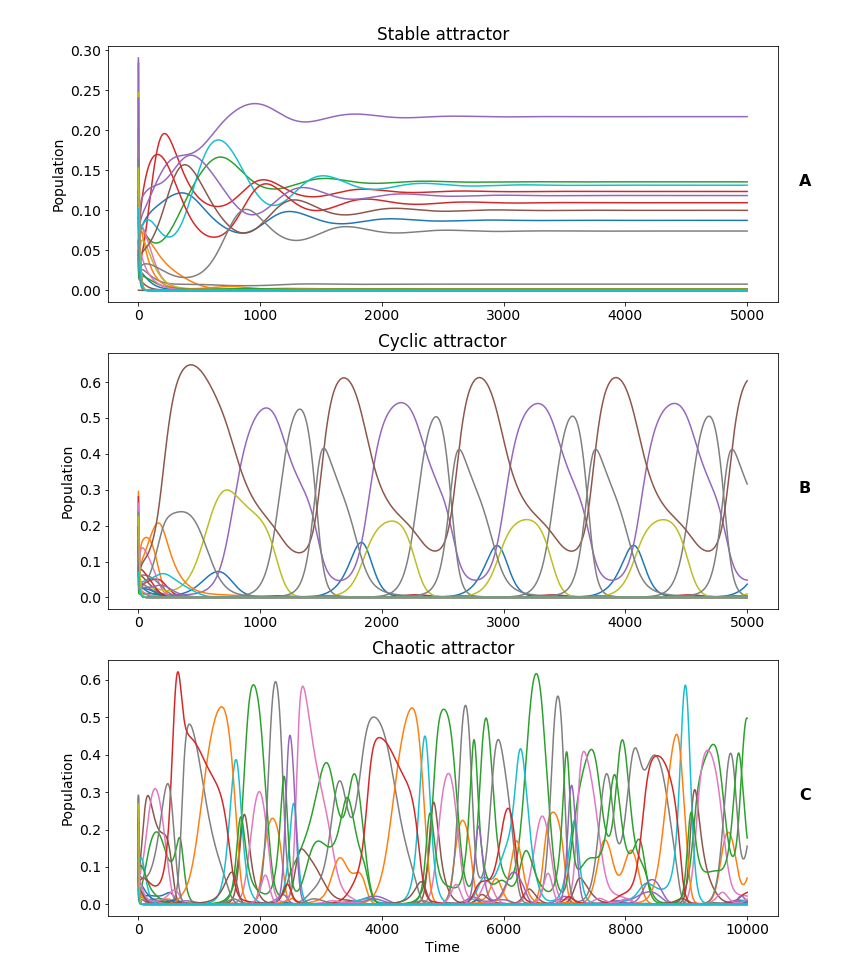}
	\end{center}
	\caption{Our family of models generates time series of the population of each species. The time series can be classified in $3$ qualitative types depending on their asymptotic behaviour: \textit{stable}, \textit{cyclic} and \textit{chaotic}. In \textbf{panel A}, the system reaches a stable attractor after a transient time. In \textbf{panel B}, a periodic attractor, with an approximate period of 1000 days, is reached after the transient time. The system in \textbf{panel C} never reaches a stable nor a cyclic attractor, but a chaotic one.}
	\label{fig:TimeSeries}
\end{figure}

Depending on the parameters and initial conditions, our model (equation \eqref{eq:SystemUnderStudy}) can have three kinds of dynamics, each of them roughly corresponding to a different kind of attractor (see figure \ref{fig:TimeSeries}). In a stable point attractor, species composition is constant. The limit cycle (and limit tori) attractors corresponds to periodically (or quasiperiodically) changing species composition. The last category are chaotic attractors, where the species composition changes irregularly within bounds and there is extreme sensitivity to initial conditions. 

% Time series generation
Our target is to estimate the probability of reaching each type of attractor under different assumptions about competition. For this, we analyzed $\numEpsilons$ values of the competition parameter $\epsilon$ (defined in section \ref{subsubsec:CompetitionParameter}), ranging from $\epsilon = -0.8$ to $\epsilon = 0.8$. The lower value was chosen to assure that the non-diagonal competition matrix elements were positive to exclude facilitation. The upper value was arbitrarily chosen to be symmetric with the lower one. For each value of the competition parameter, $\numReps$ different predation and competition matrices were drawn from the probability distributions described in section \ref{subsubsec:CompetitionParameter}. We used a Runge-Kutta solver (ode45) to simulate the model with each parameter set, using random initial conditions. A stabilizing run of $ \stabilTime $ days was executed to discard transient dynamics. Simulating for $ \simTime $ more days, we obtained a time series close to the attractor.

% Analysis
We determined the fraction of the $\numReps$ time series that were stable, cyclic or chaotic.  For our multi-species models, we found that the Gottwald - Melbourne test \citep{Gottwald2009} was the most reliable test for this classification.  Additionally, two different measures of biodiversity were applied to each simulated ecosystem: the average number of non-extinct prey species and the average biomass grouped by trophic level. We considered a species to be extinct when their population density remained below a threshold of $\bioThreshold$ $mg$ $l^{-1}$ after the stabilization run. We determined the relationship between the competition strength, the probability of each dynamical regime and the biodiversity.

The numerical experiment was repeated for food webs of different sizes, ranging from a total of $5$ to $50$ species. In our simulations, we kept a ratio of 2:3 for the number of species at the consumer and the prey level.

In the spirit of reproducible research, we made available the code used to obtain our conclusions and generate our figures \citep{Rodriguez-Sanchez-code-neuchaos}.
\section{Results}
\label{sec:Results}
\begin{figure}
	\begin{center}
		\includegraphics[width=1\columnwidth]{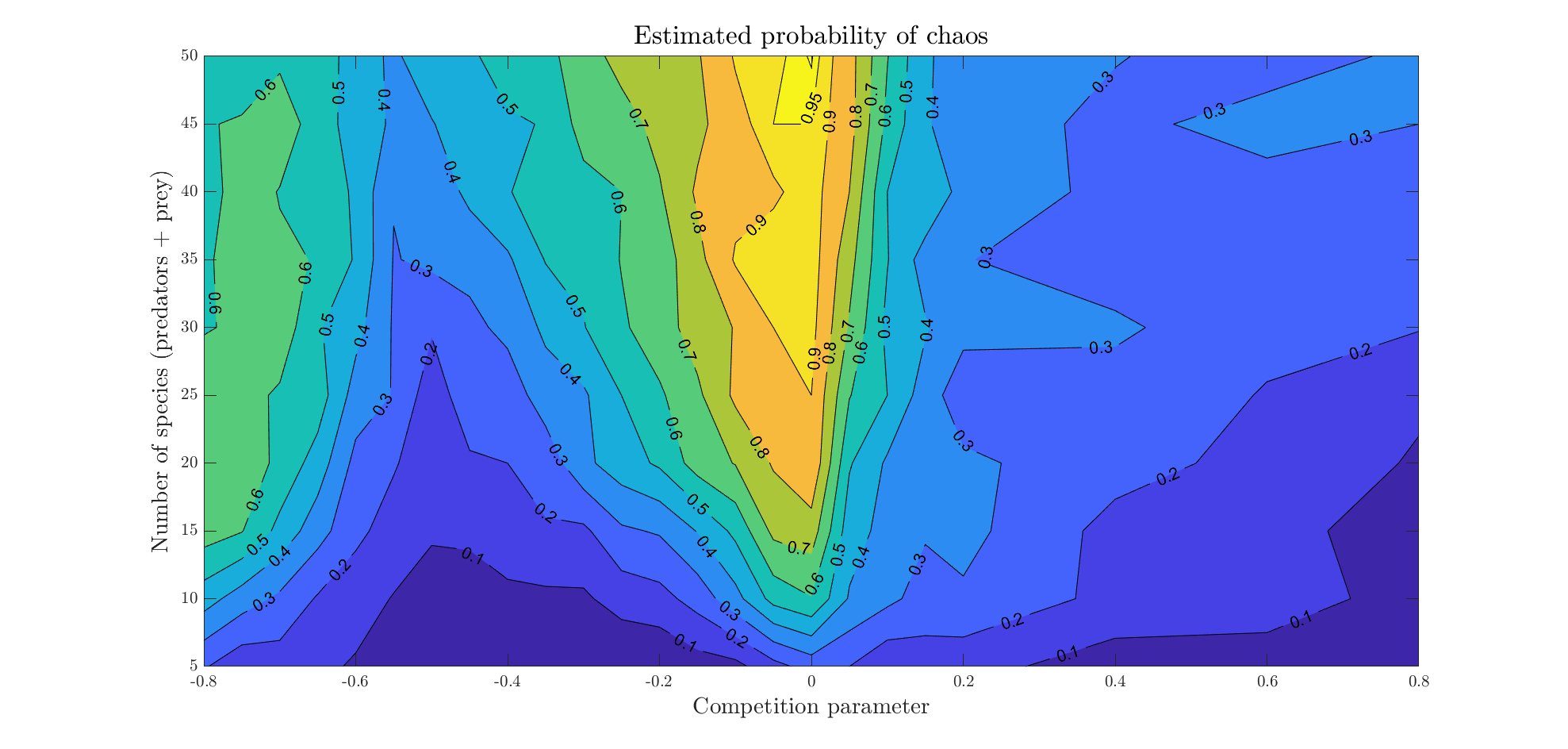}
	\end{center}
	\caption{Contour map showing the probability of chaos for various competition parameters (horizontal axis) and number of species (vertical axis). The consumers' population is fixed as $ 2/3 $ of the prey's population. Notice that chaotic attractors appear more easily (i.e., for smaller systems) the closer is the competition to neutral (i.e., $ \epsilon = 0 $).}
	\label{fig:Contour}
\end{figure}

\begin{figure}
	\begin{center}
		\includegraphics[width=1\columnwidth]{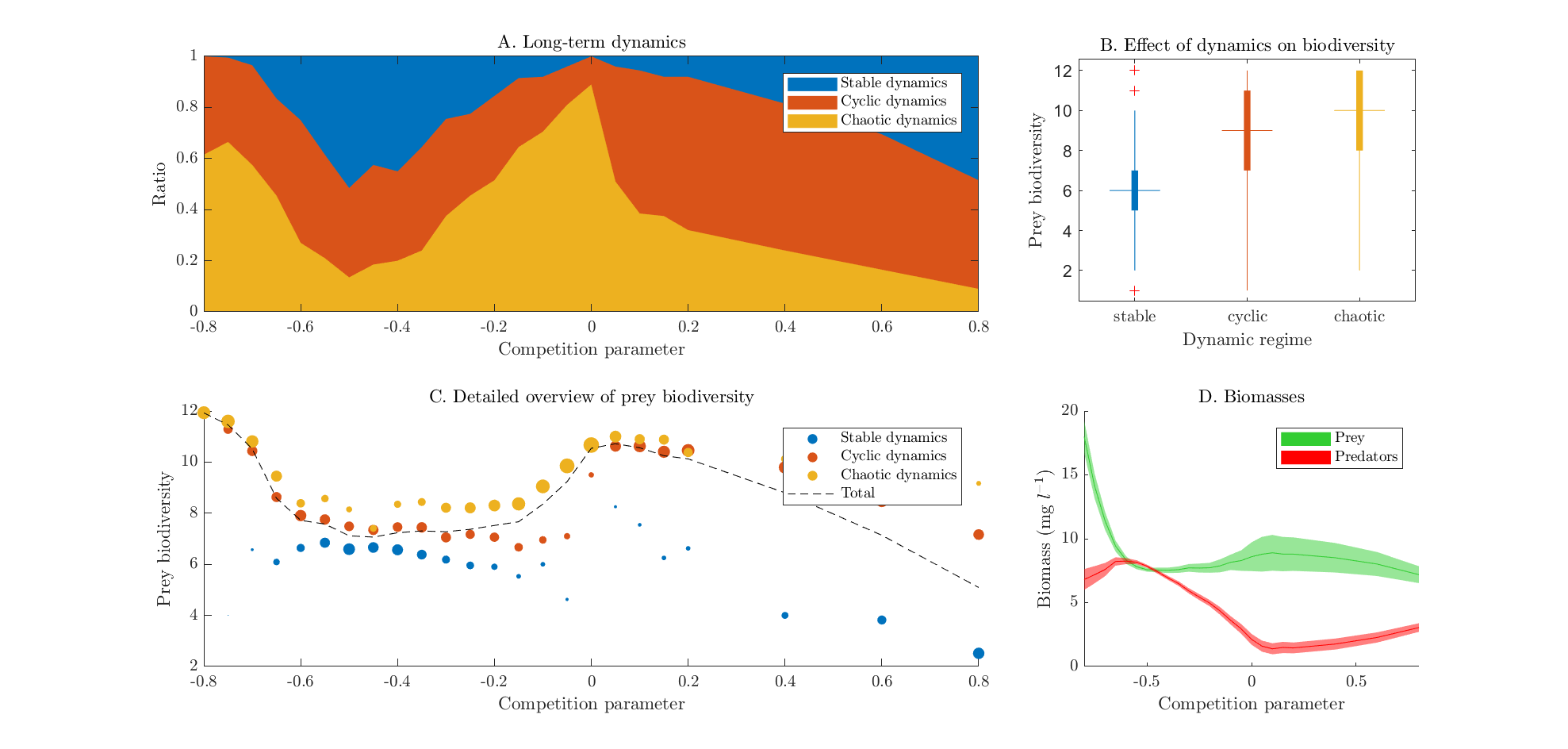}
	\end{center}
	\caption{Results for food webs with $8$ predator and $12$ prey species. Food webs of different sizes show similar results (see section \ref{subsec:GeneralResults} in Online Appendix). \textbf{Panel A}. Fraction of each dynamic regime as a function of the competition parameter ($n = \numReps$). \textbf{Panel B}. Box and whisker plot of the average number of non-extinct prey species grouped by asymptotic regime. \textbf{Panel C}. Average prey biodiversity as function of competition parameter. The dashed line shows the average number of non-extinct prey species grouped by competition parameter. The colored circles represent the average prey biodiversity of the simulations, additionally grouped by dynamical regime (stable, cyclic and chaotic). The relative size of the circles represents the ratio of simulations that led to each kind of dynamics. \textbf{Panel D}. Average biomasses grouped by trophic level vs. competition parameter. The width represents standard deviation.}
	\label{fig:Biodiversity}
\end{figure}

% Chaos
From figure \ref{fig:Contour} (see also figure \ref{fig:AllProbabilities} in the Online Appendix) we conclude that, in our model, the likelihood of chaotic dynamics reaches an optimum for near-neutral competition at the prey level. This result remains true for systems with a different number of species. The likelihood of chaos also increases with the size of the food web. This effect should not be surprising: the more dimensions the phase space has, the easier is to fulfill the requirements of the complex geometry of a chaotic attractor \citep{Strogatz1994}. Even in those higher dimensional cases, there is still a clear maximum at near-neutral competition. The probability of chaos shows another local, lower maximum for weak competition coupling, while stable solutions are very rare (figure \ref{fig:Biodiversity}.A). Possibly due to the weaker coupling we get less phase locking of the predator prey cycles in this case.

% Biodiversity
Additionally, we found a clear relationship between the probability of chaos and the biodiversity. In all our cases the diversity in systems with chaotic dynamics were highest (figures \ref{fig:Biodiversity} B,C) and the overall diversity peaked approximately at the near-neutral situation. Interestingly also the cyclic solutions were clearly much more diverse than cases with stable dynamics (figures \ref{fig:Biodiversity} B,C). In fact the difference in biodiversity of the situation with chaos and cycles was rather small (figure \ref{fig:Biodiversity}.C). This conclusion remains true for food webs of different sizes (figure \ref{fig:BiodBoxAndWhisker} in the Online Appendix). From figure \ref{fig:Biodiversity}.D, we see that the prey biomass remains relatively stable for the whole range of competition parameters, with the exception of weak interspecific competition, where it reaches a maximum. The predator biomass grows almost linearly as the competition moves leftwards, from near-neutral to strong intraspecific, while the prey biomass remains constant. We think this can be understood from the effect of niche complementarity which causes effectively an increase in the total prey biomass. Like in a two-species model this increase in prey biomass results in an increase of predator biomass only (cf. \citet{Rosenzweig1963}).

\section{Discussion}
\label{sec:Discussion}
\begin{figure}
	\begin{center}
		\includegraphics[width=0.5\columnwidth]{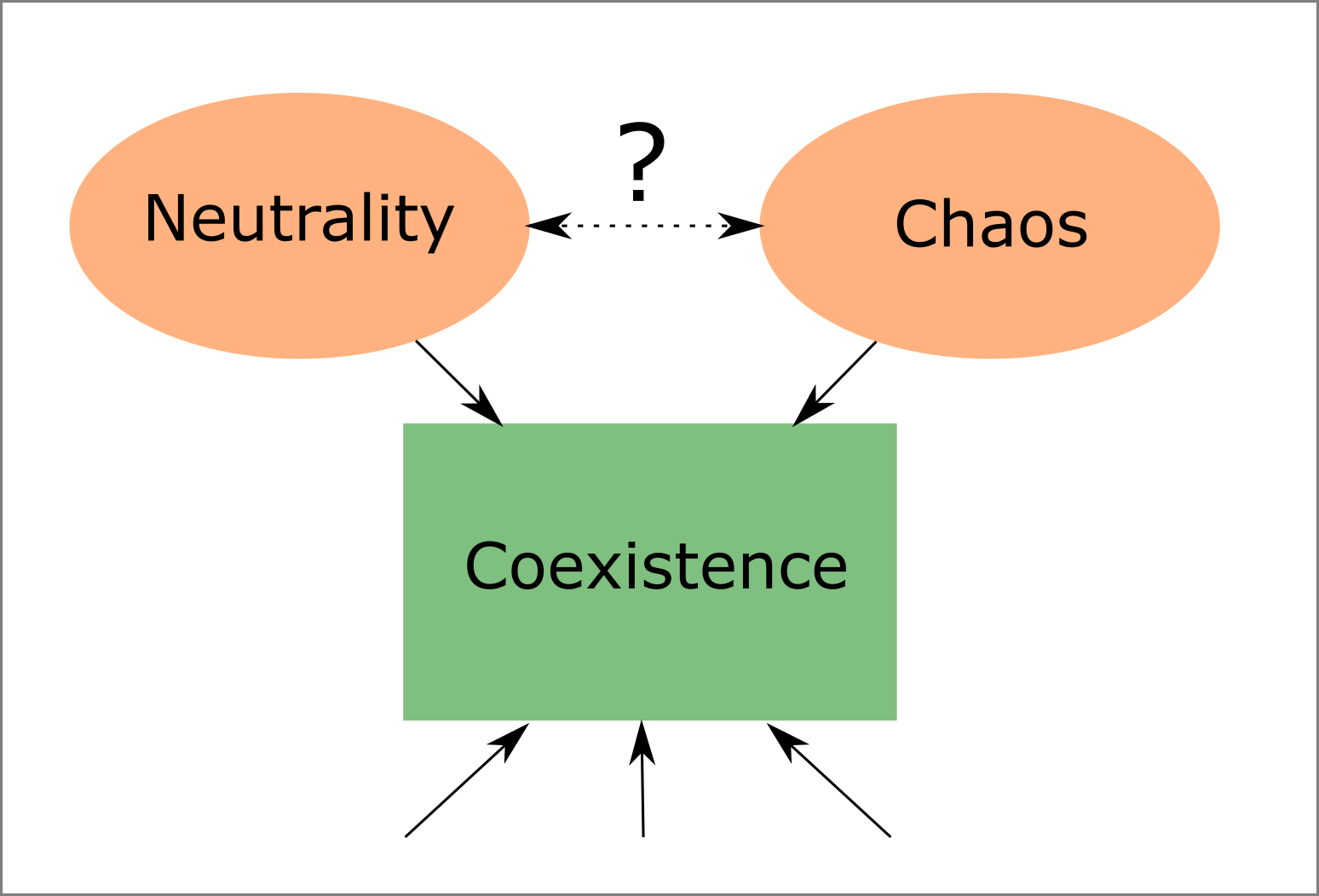}
	\end{center}
	\caption{In our model, neutrality and chaos are not independent explanations of coexistence.}
	\label{fig:GapInKnowledge}
\end{figure}

% Our claims
We find that competition close to neutrality significantly increases the chances of chaotic behaviour. This peak in chaotic dynamics indeed coincided with a peak in biodiversity. These observations, suggest that the hypothesis of non-equilibrium \citep{Huisman1999} and Hubbell's hypothesis of neutrality are not completely independent (see figure \ref{fig:GapInKnowledge}). Additionally we observed that also non-chaotic periodic dynamics  led to a higher biodiversity than stable equilibria.

%% Limitations
Our result seemed to be robust against changes in the number of species. However, the probability of chaotic dynamics is dependent on the model details and all parameters \citep{Dakos2009b}. We explored the variation away from neutrality only by changing the competition strength. We didn't use Allee effects, nor noise, nor species-specific carrying capacities, and the functional form of each term has been chosen to account for satiation and saturation in the simplest possible ways. This opens the door to perform similar analyses in the future using more sophisticated models.

Both the competition and predation parameter sets were drawn from probability distributions. The interactions in our system can be interpreted as a weighted network with a high connectivity. In nature, trophic networks tend to show modular structure with various clusters \citep{Thebault2010}. Our simplified model could be interpreted as representing one of those densely connected modules. Moreover, while in the present paper our random parameters were drawn independently, the competition matrix can be chosen in a more advanced way (for instance, accounting for rock-paper-scissors competition). Studying the effect of different physiological scenarios (in the sense of \citet{Huisman2001}, that is, constrains between the parameters) on the probabilities of chaos could be a continuation to this paper.

% Chaos detection
Due to the large number of simulations made, we had to rely on automatic methods for detecting chaos. Automatic detection of chaos by numerical methods has fundamental limitations, especially for high dimensional systems like ours. Most of them can be boiled down to the fact that, in general, numerical methods cannot distinguish robustly between long, complicated transients and genuine chaos. Our motivation to choose the Gottwald - Melbourne test \citep{Gottwald2009} was threefold: it discriminates between stable, cyclic and chaotic, it escalates easily to systems of higher dimensions, its computation is fast and it performs better than any other method we tried when compared to the visual inspection of the time series. Although we cannot exclude that we misinterpreted some of the generated time series due to long transients, we don't think this affected the overall patterns, as they were very robust in all our simulations.

% Concluding remark
Our results suggest a fundamentally new way in which near-neutrality may promote biodiversity. In addition to weakening the forces of competitive exclusion \citep{Scheffer2018}, our analyses reveal that near neutrality may boost the chances for chaotic dynamics. As chaos and cycles may facilitate super-saturated co-existence, our findings point to a potentially widespread mechanism of maintaining biodiversity.

\section{Acknowledgments}
\label{sec:Acknowledgments}
The preliminary analysis of this model was performed using GRIND for Matlab (\url{http://www.sparcs-center.org/grind}). Additionally, we thank Tobias Oertel-Jäger, Sebastian Wieczorek, Peter Ashwin, Jeroen Lamb, Martin Rasmussen, Cristina Sargent, Jelle Lever, Moussa N'Dour, Iñaki Úcar, César Rodríguez and Sebastian Bathiany for their useful comments and suggestions. 

This work was supported by funding from the European Union's \textit{Horizon 2020} research and innovation programme for the \textit{ITN CRITICS} under Grant Agreement Number 643073.

%% Appendix
\clearpage
\appendix
\setcounter{equation}{0}
\setcounter{figure}{0}
\renewcommand{\theequation}{A.\arabic{equation}}
\renewcommand\thefigure{A.\arabic{figure}}

\section{Online appendix}
\label{sec:Appendix}

This is the Online Appendix for the paper:

\begin{center}
Rodríguez-Sánchez P, van Nes EH, Scheffer M. \textit{Neutral competition boosts chaos in food webs}.
\end{center}

\newpage

\subsection{Results for food webs of different sizes}
\label{subsec:GeneralResults}
In the main body of the paper we focused our attention in families of food webs consisting of $12$ prey and $8$ predator species. In this section we show the results of the same analysis for food webs of different sizes.

\subsubsection{Probability of chaos grouped by number of species}
\label{subsubsec:AllProbabilities}
\begin{figure}[H]
	\begin{center}
		\includegraphics[width=1\columnwidth]{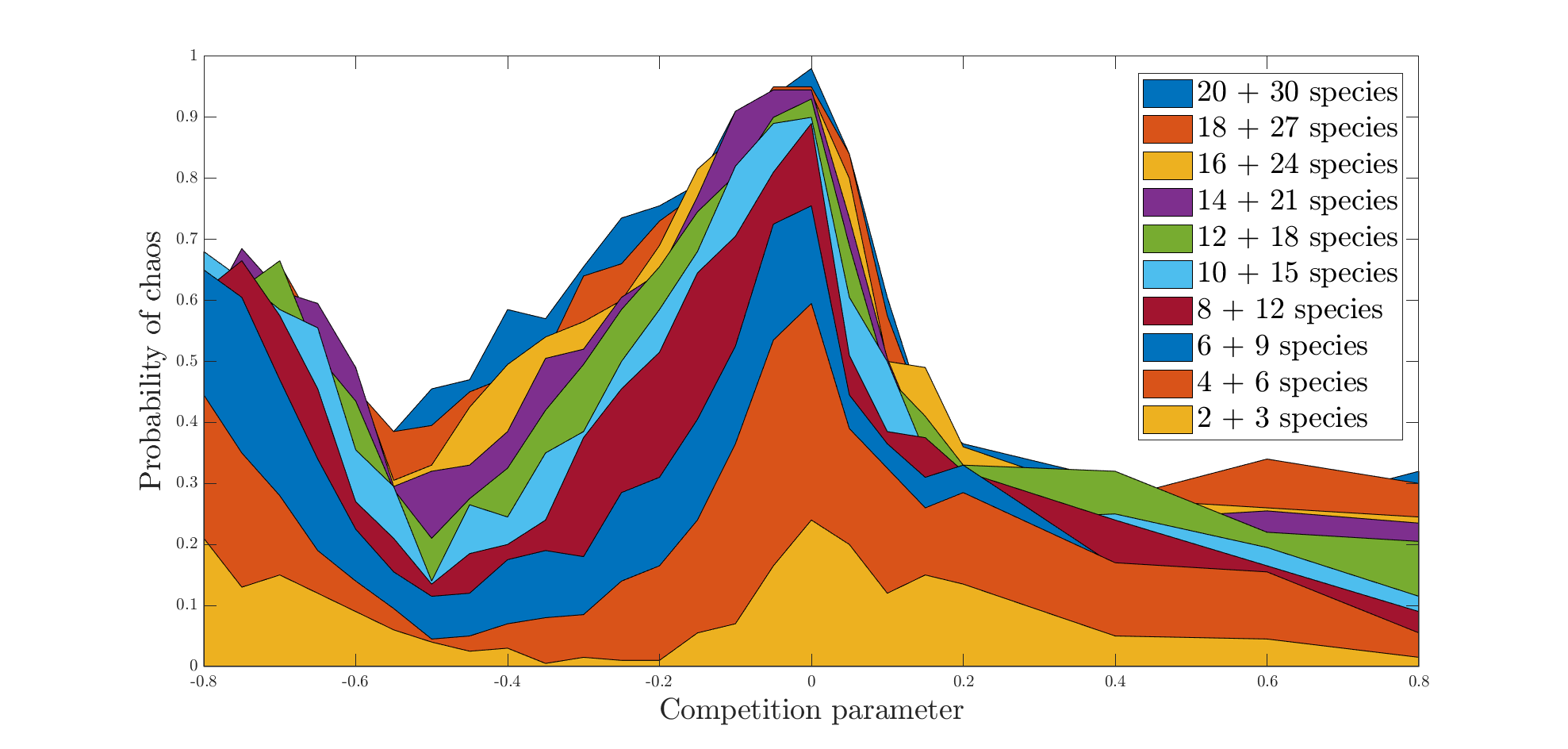}
	\end{center}
	\caption{Probabilities of chaos vs. competition parameter for the whole set of simulations. The competition parameter $\epsilon$ is on the horizontal axis. The estimated probability of chaos is represented on the vertical one. Each panel corresponds to an ecosystem with a different number of interacting species. The exact number is shown in each box, as number of predator + number of prey species.}
	\label{fig:AllProbabilities}
\end{figure}

\subsubsection{Probability of each dynamical regime}
\label{subsubsec:DynamicalRegimes}
\begin{figure}[H]
	\begin{center}
		\includegraphics[width=1\columnwidth]{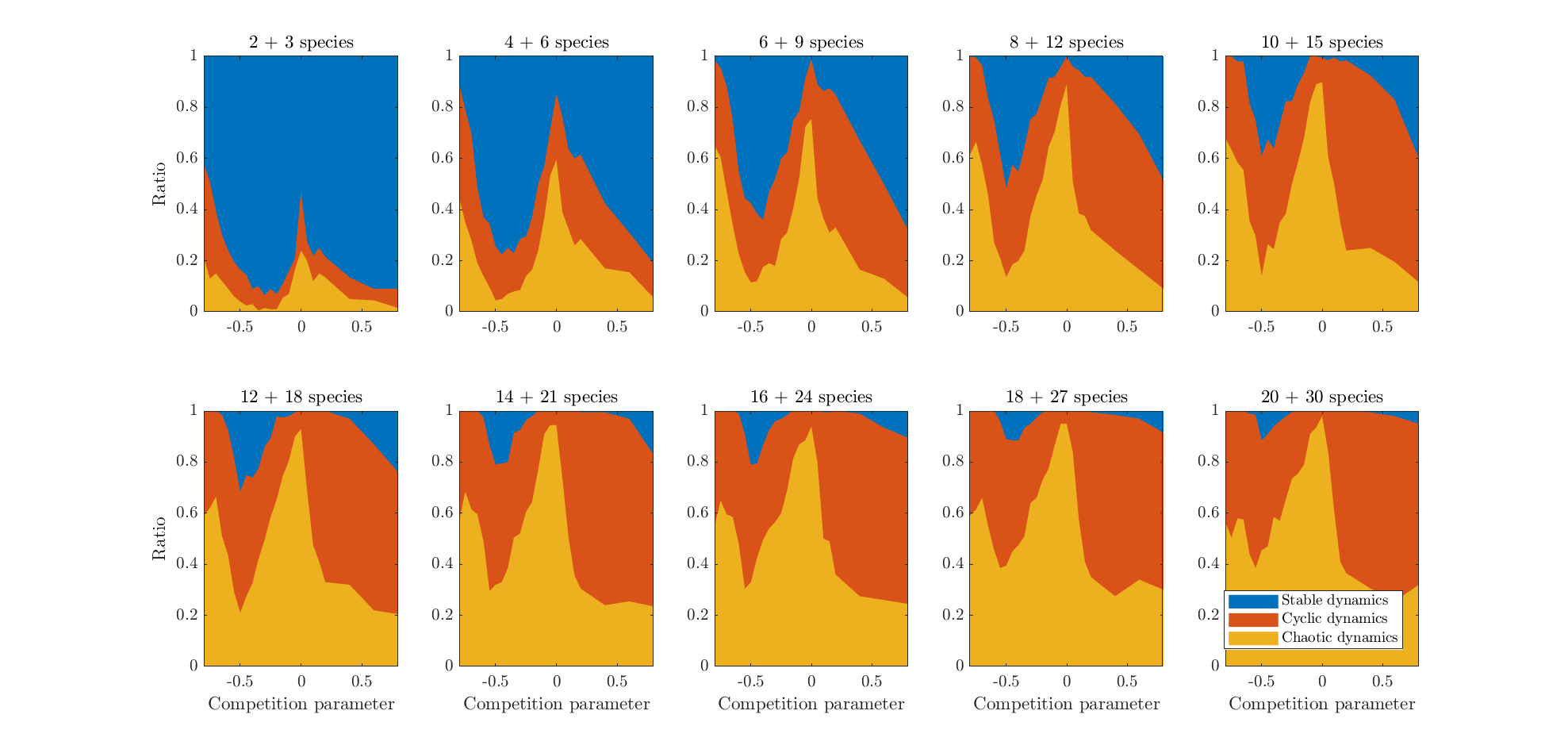}
	\end{center}
	\caption{Ratio of each dynamical regime vs. competition parameter for the whole set of simulations. The competition parameter $\epsilon$ is on the horizontal axis. The system size is shown in each box, as number of predator + number of prey species.}
	\label{fig:DynamicalRegimes}
\end{figure}

\subsubsection{Biodiversity measurements}
\label{subsubsec:BiodiversityFigs}

For each simulation, a biodiversity index was estimated as the number of prey species whose population was higher than a minimum threshold of $\bioThreshold$ $mg$ $l^{-1}$, averaged respective to time.

\begin{figure}[H]
	\begin{center}
		\includegraphics[width=1\columnwidth]{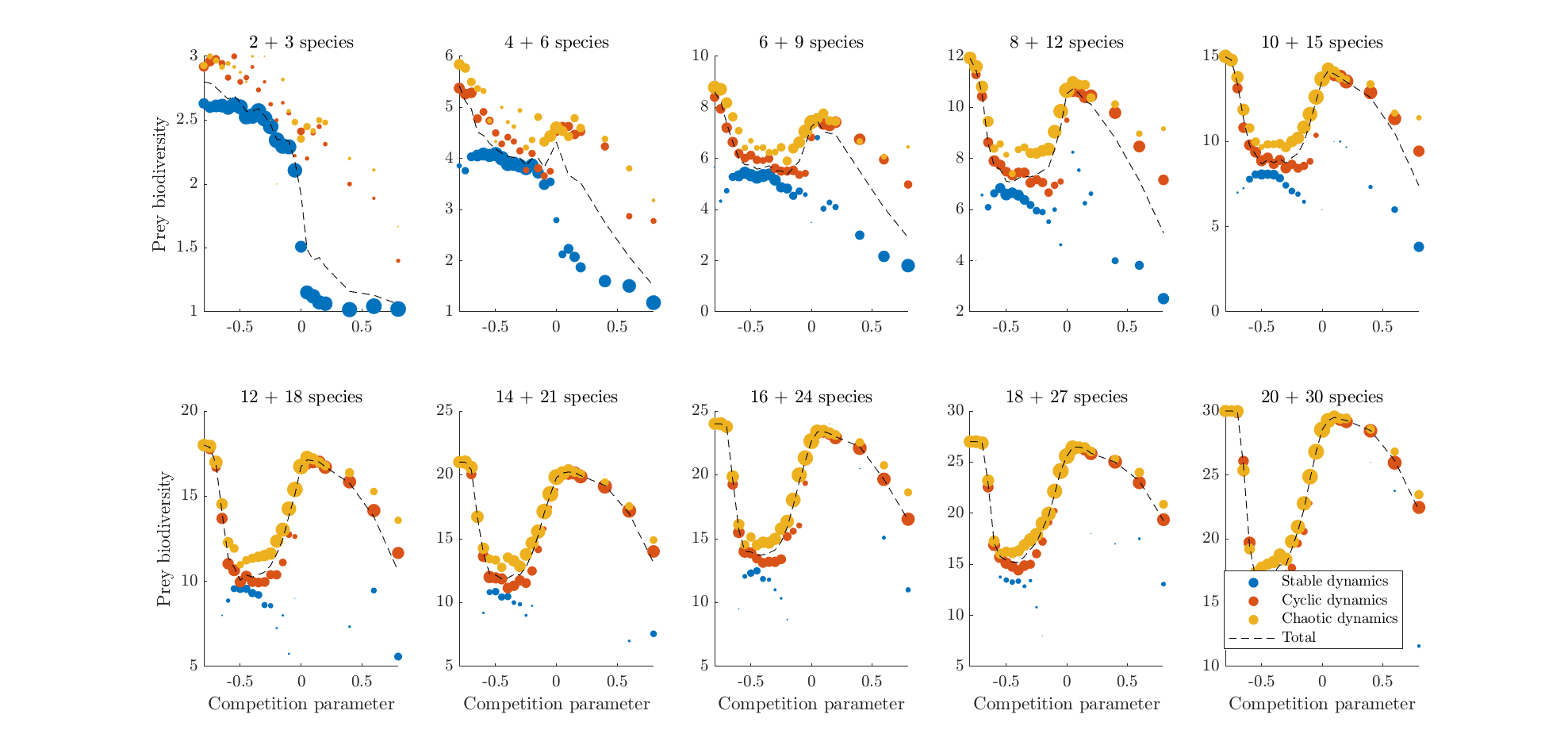}
	\end{center}
	\caption{Average prey biodiversity vs. competition parameter. Each panel shows a food network of a different size. For each value of the competition parameter, 200 randomly drawn ecosystems were simulated. The dashed line shows the average number of prey species of these 200 simulations. The yellow circles represent the average prey biodiversity of those simulations who had chaotic dynamics. The red and blue circles represent the same for, respectively, cyclic and stable dynamics. The relative area of the circles represents the ratio of each kind of dynamics.}
	\label{fig:BiodSplitByChaos}
\end{figure}

\begin{figure}[H]
	\begin{center}
		\includegraphics[width=1\columnwidth]{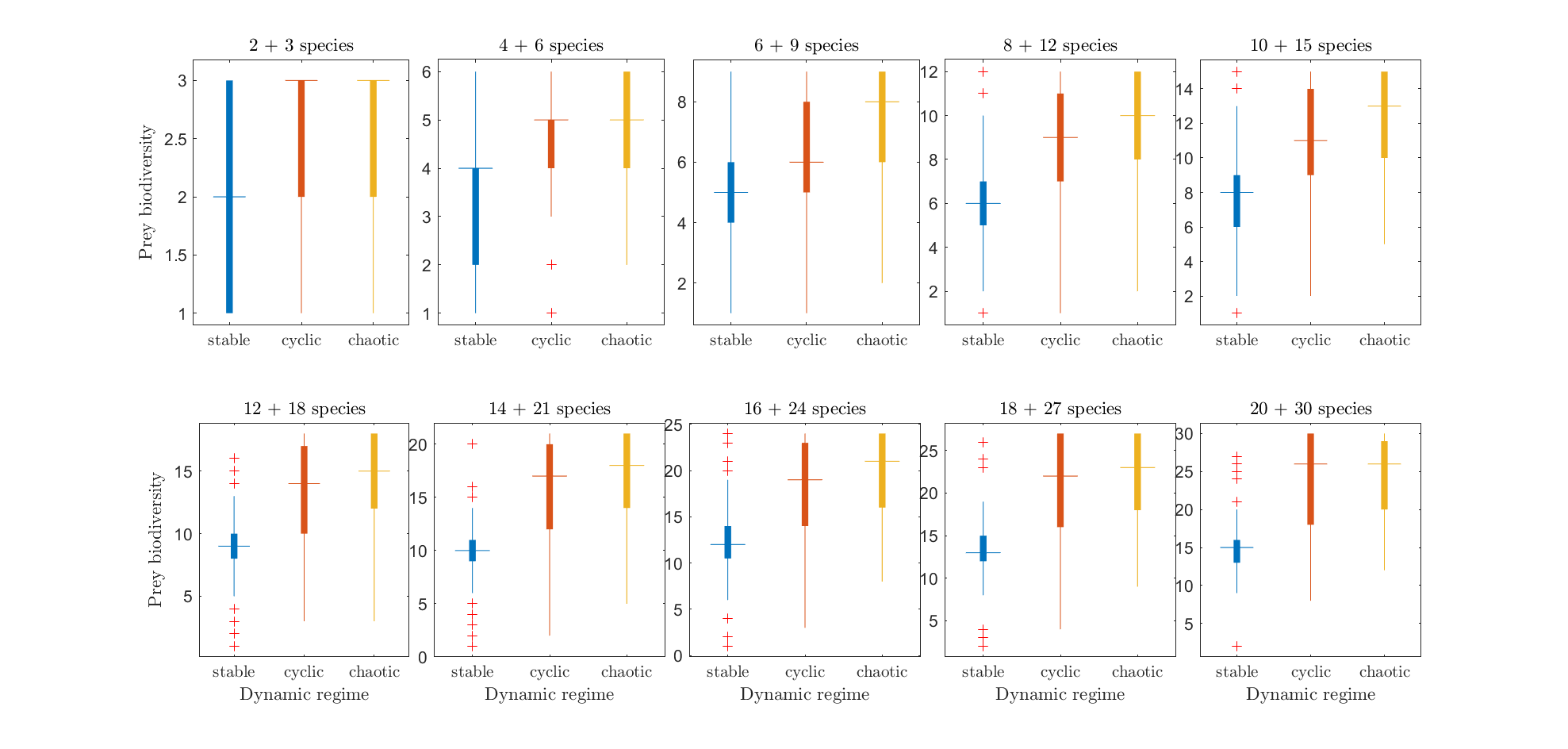}
	\end{center}
	\caption{Box and whisker plot of the prey biodiversity, after being classified as stable, cyclic or chaotic.}
	\label{fig:BiodBoxAndWhisker}
\end{figure}

\begin{figure}[H]
	\begin{center}
		\includegraphics[width=1\columnwidth]{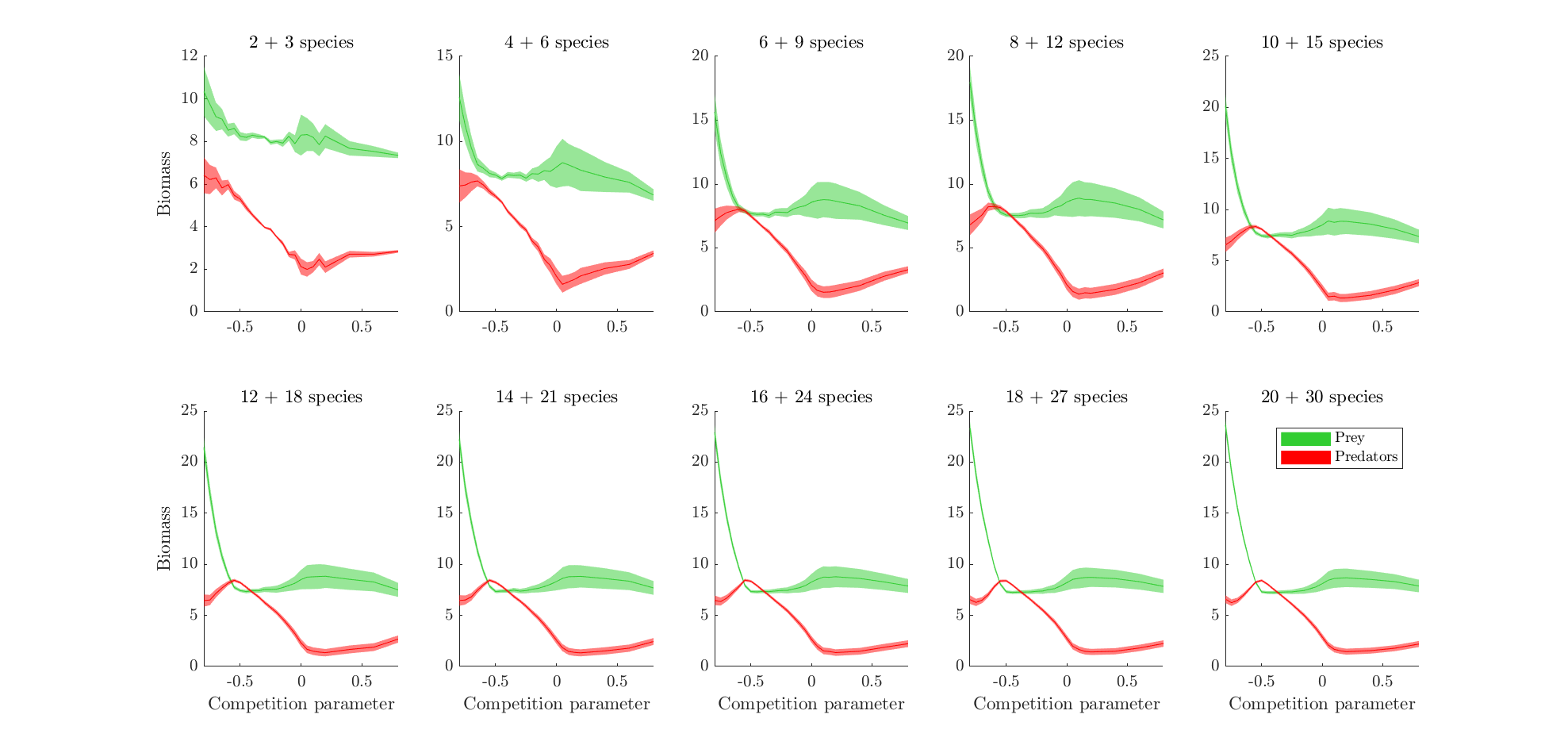}
	\end{center}
	\caption{Average biomasses grouped by trophic level vs. competition parameter. The width represents standard deviation.}
	\label{fig:Biomass}
\end{figure}

%% Bibliography
\clearpage
\bibliography{./bib/library}

\end{document}